\documentclass[referee]{cjaa}
\input{epsf.sty}
\usepackage{graphicx}
\input{psfig.sty}

\begin{document}
   \title{Elliptical Galaxies with Emission Lines from the Sloan Digital Sky Survey}

   \author{Ying-He Zhao, Qiu-Sheng Gu\mailto{}, Zhi-Xin Peng, Lei Shi, Xin-Lian Luo \and Qiu-He Peng}

   \institute{Department of Astronomy, Nanjing University, Nanjing 210093, China\\
   \email{qsgu@nju.edu.cn}
}

   \date{Received ~~~~~~~~; accepted ~~~~~~~~}

   \abstract{
   We present the results of 11 elliptical galaxies with strong nebular emission
   lines during our study of star formation history along the Hubble sequence.
   After removing the dilution from the underlying old stellar populations
   by use of stellar population synthesis model, we derive the accurate fluxes
   of all emission lines for these objects, which are later classified with emission
   line ratios into one Seyfert 2, six LINERs and four HII galaxies. We also
   identify one HII galaxy (A1216+04) as a hitherto unknown Wolf-Rayet galaxy from the presence
   of the Wolf-Rayet broad bump at 4650 \AA. We propose that the
   star-forming activities in elliptical galaxies are triggered by either galaxy-galaxy
   interaction or the merging of a small satellite/a massive star cluster, as
   already suggested by recent numerical simulations.
   }
\authorrunning{Y. H. Zhao, Q. S. Gu \& Z. X. Peng et al.}
\titlerunning{Elliptical galaxies with emission lines}

 \maketitle


\section{Introduction}

   The star formation history of elliptical galaxies carries an
   important information for their formation and evolution and has
   invaluable constraints for the cosmological models (Eggen,
   Lynden-Bell \& Sandage 1962). In the conventional view, an
   elliptical galaxy was thought to be a simple stellar system
   with old stellar population formed in a single star forming
   activity at very early stage of the galaxy forming history,
   quiescent and passively evolving ever since, and very few new
   stars are made in the past 1$-$ 2 Gyrs (Searle, Sargent \&
   Bagnuolo 1973; Larson 1975). However, in recent decades a rival
   view has been proposed based on the hierarchical clustering
   models (White \& Rees 1978; Kauffmann, White \& Guiderdoni
   1993), which requires that the most massive objects form at
   late time via merging smaller subunits. In this scenario,
   massive ellipticals are formed through merging of two spiral
   galaxies, as have also been shown by numerical simulations
   (see the recent review by de Freitas Pacheco, Michard \&
   Mohayaee 2003; Toomre \& Toomre 1972; Barnes 1992; Barnes \&
   Hernquist 1992; Bendo \& Barnes 2000).

   Recently Fukugita et al. (2004) have reported the discovery of
   active star-forming activities in the field elliptical galaxies
   from the Sloan Digital Sky Survey (SDSS). They found that the
   percentage of such star-forming elliptical galaxies is a few
   tenths of a percent in their sample and suggested that these
   star-forming ellipticals could be the progenitors of E$+$A
   galaxies, which are devoid of nebular emission lines but with
   very strong Balmer absorption lines superimposed with an old
   stellar population, and are interpreted as being the
   post-starburst ended within the last 1.5 $-$ 2 Gyr (Couch \&
   Shaples 1987; Barger et al. 1996). In this paper, we will
   present more examples of such star-forming elliptical galaxies
   which show the unambiguous evidence of young star-forming
   activities during our study of star formation history along the
   Hubble sequence.

   This paper is organized as follows: we give the sample of 11
   ellipticals with strong emission lines in Section 2. In Section
   3, we present the results of stellar population synthesis,  use
   the standard BPT diagrams to classify the ionizing mechanism for
   these 11 ellipticals, and study radial profiles, color
   distribution and star forming activities for three HII
   galaxies (A1212+06, A1216+04 and CGCG13-83). Finally we
   discuss in Section 4 and present our conclusions in Section 5.
   Where appropriate we adopt a Hubble constant of H$_0$ = 75 km
   s$^{-1}$ Mpc$^{-1}$,  $\Omega_M$ = 0.3 and $\Omega_\Lambda$ =~
   0.7.


\section{The Data}

 The Sloan Digital Sky Survey (SDSS) is the most ambitious
 astronomical (both photometric and spectroscopic) survey project
 which has ever been undertaken (Gunn et al. 1998; Blanton et al. 2003).

 Recently, we cross-correlate the SDSS DR2 spectroscopic archive
 dataset with the Third Reference Catalogue of Bright Galaxies
 (RC3; de Vaucouleurs et al. 1991) by positional match with
 accuracy of $\sim$ 10 arcsec, and derive a sample of 1027
 galaxies with both morphological classification and spectroscopic
 information, among of which 48 sources are cataloged as
 Elliptical galaxies in RC3 with mean numerical index (T) of
 either -4 or -5. Our main target is to study the star formation
 activity along the Hubble sequence, which will be presented by
 Shi et al. (2005).

 In this paper we show that among these 48 elliptical galaxies, 11
 objects clearly present strong nebular emission lines with
 H$\alpha$ equivalent width (EW) larger than 2\AA, which are not
 classified as AGNs before either in the catalog of {\it Quasars
 and Active Galactic Nuclei} (Veron-Cetty \& Veron 2003, 11th Ed.)
 or in any literature. The basic properties of these 11
 ellipticals are summarized in Table 1, where we give the galaxy
 name, coordinate, redshift, distance, absolute blue magnitude and
 morphological type, respectively.

 \begin{table}[h]
\caption{Parameters of 11 elliptical galaxies with emission lines from SDSS}
\begin{center}
\begin{tabular}{lcccccc}
\hline\noalign{\smallskip}
Name &RA (2000)&DEC (2000)&Redshift& Distance (Mpc)& M$_B$ & Morphology \\
\hline\noalign{\smallskip}
NGC\ 426    & 01h12m48.6s & -00d17m24.6s & 0.0173 & 69.2 &-20.4 & E+\\
NGC\ 677    & 01h49m14.0s & +13d03m19.1s & 0.0170 & 68.0 &-21.0 & E0\\
IC\ 225     & 02h26m28.2s & +01d09m37.9s & 0.0051 & 20.4 &-17.1 & E0\\
CGCG 13-83 & 12h08m23.5s & +00d06m36.9s & 0.0408 & 163.2 &-21.2 & E0\\
NGC4187   & 12h13m29.2s & +50d44m29.3s & 0.0305 & 122.0 &-21.4 & E0\\
A1212+06  & 12h15m18.3s & +05d45m39.4s & 0.0067 & 26.8 &-17.3 & E0\\
A1216+04  & 12h19m09.8s & +03d51m23.3s & 0.0051 & 20.4 &-17.1 & E0\\
NGC4581   & 12h38m05.1s & +01d28m39.9s & 0.0062 & 24.8 &-18.8 & E+\\
NGC5216   & 13h32m06.9s & +62d42m02.4s & 0.0979 & 39.2 &-19.5 & E0\\
IC\ 989   & 14h14m51.3s & +03d07m51.3s & 0.0253 & 101.2 &-21.3 & E0\\
NGC5846   & 15h06m29.1s & +01d36m20.9s & 0.0057 & 22.8 &-20.9 & E0\\
\hline\noalign{\smallskip}
\end{tabular}
\end{center}
\end{table}

\section{Results}

 During the study of emission-line spectra, an unavoidable issue
 is the contamination from the underlying old stellar population,
 especially to the Balmer lines. The standard method is to use the
 stellar population synthesis model to fit the observed spectrum.
 In this paper, we use the same stellar population synthesis code,
 {\scriptsize STARLIGHT version 2.0}, (Cid Fernandes et al. 2004)
 to study 11 elliptical galaxies. The routine searches for the
 linear combination of 45 Simple Stellar Populations (SSP) from
 the recent stellar population model of Bruzual \& Charlot (2003)
 for 3 metallicities (0.2 Z$\odot$, Z$\odot$, and 2.5 Z$\odot$),
 for each metallicity we select 15 different age components,
 ranged as 0.001, 0.003, 0.005, 0.01, 0.025, 0.04, 0.10, 0.29,
 0.64, 0.90, 1.4, 2.5, 5.0, 11 and 13 Gyr. The match between model
 and observed spectrum is evaluated by the ruler of $\chi$ square
 minimum, and the search for the best parameters is carried out
 with a simulated annealing method, which consists of a series of
 10$^7$ likelihood-guided Metropolis tours through the parameter
 space (see also Cid Fernandes et al. 2005).

 After subtracting the best-fit model from the observed one, we
 derive the pure emission-line spectrum, from which we could
 measure the accurate fluxes for all emission lines with the IRAF
 \footnote{IRAF is distributed by the National Optical Astronomy
 Observatories, which is operated by Associated of Universities
 for Research in Astronomy, Inc., under cooperative agreement with
 the National Science Foundation.} {\it specfit} or {\it
 onedspec.splot} software, the results are presented in Table 2.
 The line flux errors are typically around 5\%. It is well known
 that we can distinguish narrow-line AGNs from normal star-forming
 galaxies by using emission line flux ratios, the so-called BPT
 diagrams (Baldwin, Phillips \& Terlevich 1981; Veilleux \&
 Osterbrock 1987). With our measurement of emission lines from the
 pure emission-line spectra, we classify these 11 ellipticals into
 three types: one Seyfert 2, six LINERs and four HII galaxies(see
 Figure 1). We find that three HII galaxies (A1212+06, A1216+04
 and IC 225) are in fact dwarf galaxies with absolute B magnitude
 (M$_B$) smaller than -18, as shown in Table 1.

\begin{figure}
\begin{center}
\epsscale{1.0}{1.0} \plotone{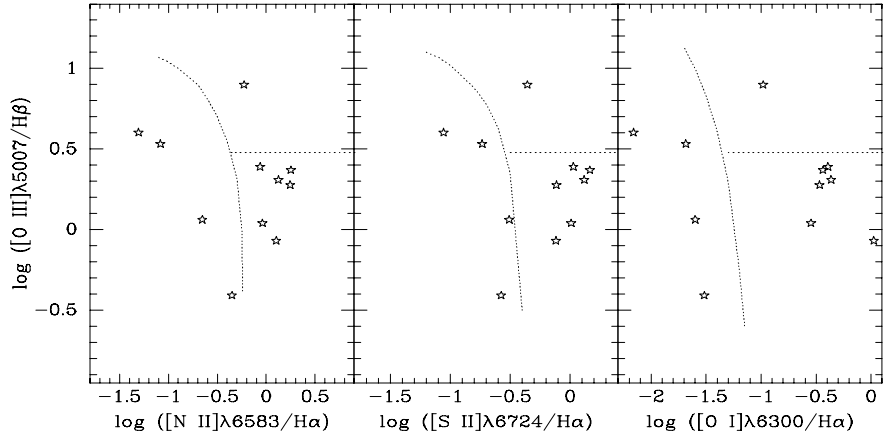} \caption{Diagnostic
diagrams. \emph{left.} log([N II]$\lambda$6583/H$\alpha$) vs.
log([O III]$\lambda$5007/H$\beta$); \emph{middle.} log([S
II]$\lambda$6724/H$\alpha$) vs. log([O
III]$\lambda$5007/H$\beta$); \emph{right.} log([O
I]$\lambda$6300/H$\alpha$) vs. log([O III]$\lambda$5007/H$\beta$).
The dotted lines divided narrow-line AGNs from starburst galaxies
are taken from Veilleux \& Osterbrock (1987).}
\end{center}
\end{figure}

\begin{table}[h]
\caption{Emission Lines Properties and Classifications for 11
Elliptical Galaxies}
\begin{center}
\begin{tabular}{lccccccccccc}
\hline\noalign{\smallskip}
Galaxy&&&&Flux$^a$&&&&Type\\
\hline\noalign{\smallskip} &H$\beta$&[O III]$\lambda4959$&[O
III]$\lambda$5007&[O I]$\lambda$6300&
H$\alpha$& [N II]$\lambda$6583& [S II]$\lambda$6724& \\
\hline\noalign{\smallskip}
A1212+06& 3404.6 & 3809.3& 11607.2 & 207.1 &10013.9 &829.0 & 1846.7 & HII\\
A1216+04& 4224.0 & 5648.2&16918.5 & 127.8 & 18514.3 &914.1 & 1628.5 & HII\\
CGCG13-83&239.4 & 35.4 & 93.5 & 31.0 & 1017.8 &454.3 & 270.2 & HII\\
IC 225&862.6 & 331.5& 994.6& 90.2 &3262.3 & 720.9 &1008.4 & HII\\
IC 989&171.4 & 97.9 & 349.0 & 302.5& 698.8 &931.6 & 914.1 & LINER\\
NGC4187& 246.6 & 176.2 & 576.9 & 351.7&963.0 &1720.0 & 1404.7 & LINER\\
NGC 426& 579.9 & 297.4 & 1099.2 & 910.8 &2681.4 &4708.7 & 2056.7& LINER\\
NGC4581& 909.0 & 2354.2 & 7203.8 & 405.8 &3893.6 &2310.0 & 1711.5 & Seyfert\\
NGC5216& 124.7 & 67.3 &306.0 & 238.3 & 588.7 & 509.8 & 626.3 & LINER\\
NGC5846&302.8 & 160.6 & 258.4 & 695.8&658.6 &833.2 & 501.3 & LINER \\
NGC 677& 263.1 & 71.8 & 288.4 & 303.3&1069.2&979.5 & 1088.7 & LINER\\
\hline\noalign{\smallskip}
\end{tabular}
\end{center}
$^{a}$ fluxes are in units of 10$^{-17}$ erg s$^{-1}$ cm$^{-2}$
\end{table}

 It will be easy to understand Seyfert and LINER activities in
 elliptical galaxies, such as the host galaxies of the most
 powerful AGNs (quasars) are usually the massive elliptical
 galaxies (Floyd et al. 2004), that is because the recent
 discovery of a tight relation between bulge velocity dispersion
 and black hole mass, which strongly suggests that massive black
 holes are ubiquitous and scale with the bulge masses (Ho \&
 Kormendy 2000; Tremaine et al. 2002). While the pure star-forming
 activities in elliptical galaxies are especially rare, the nature
 and triggering mechanism are still open questions. In the
 following part of this paper, we will concentrate on the detailed
 study of three star-forming galaxies (A1212+06, A1216+04 and
 CGCG 13-83). For the fourth one (IC 225), Gu et al. (2005)
 discovered it is a compact dwarf elliptical galaxy with a
 peculiar blue core which contains two distinct nuclei, separated
 by 1.4 arcseconds. The off-nucleated core could be a dwarf galaxy
 or a halo cluster, swallowed by IC 225 and thus triggered the
 starburst activity in IC 225.

 In Figure 2, we show the false-color RGB images for 4
 star-forming elliptical galaxies, which combine information from
 g-, r- and i-band SDSS images by using the algorithm given by
 Lupton et al. (2004). In Figure 3, we plot their SDSS optical
 spectra, where the insets in three panels (a, c, d) are the
 enlarged views of the higher-order Balmer absorption lines in the
 wavelength of 3750 -- 4150 \AA. The inset in panel (b) shows
 enlargement on the Wolf-Rayet blue bump around 4650 \AA.

\begin{figure}
\begin{center}
\epsscale{1.0}{1.0} \plotone{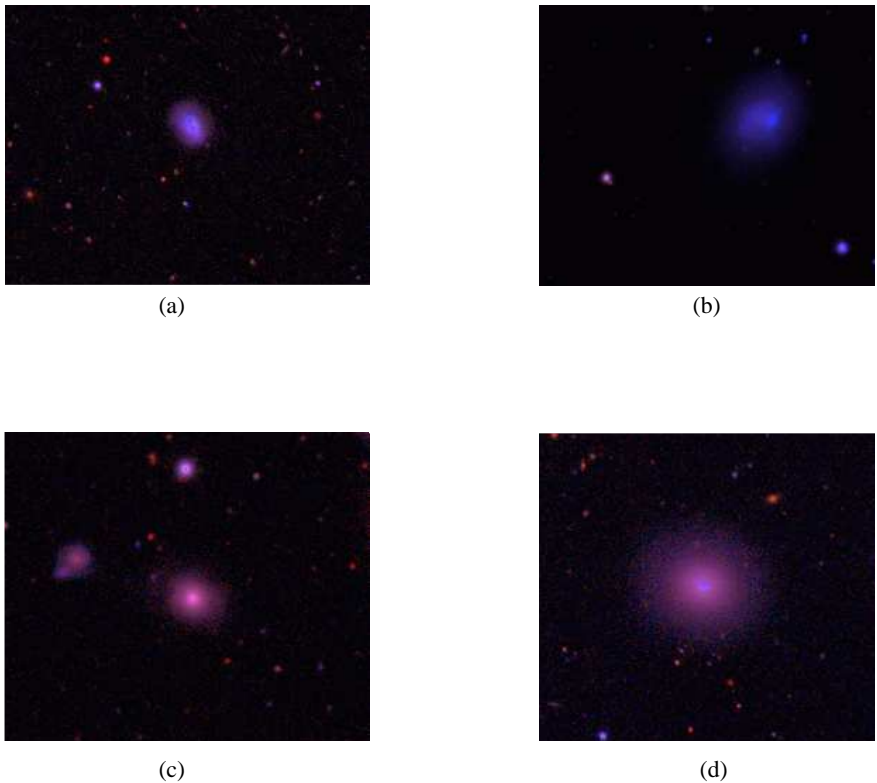} \caption{The false-color
RGB images of 4 starburst galaxies, which are obtained by using
three (r, g, i) SDSS images with the method proposed by Lupton et
al (2004). (a) A1212+06, $3'.5 \times 2'.5$; (b) A1216+04, $3'.2
\times 2'.5$; (c) CGCG 13-83, $3'.5 \times 2'.8$; (d) IC 225,
$3'.2 \times 2'.8$.}
\end{center}
\end{figure}

\begin{figure}
\begin{center}
\epsscale{1.0}{1.0} \plotone{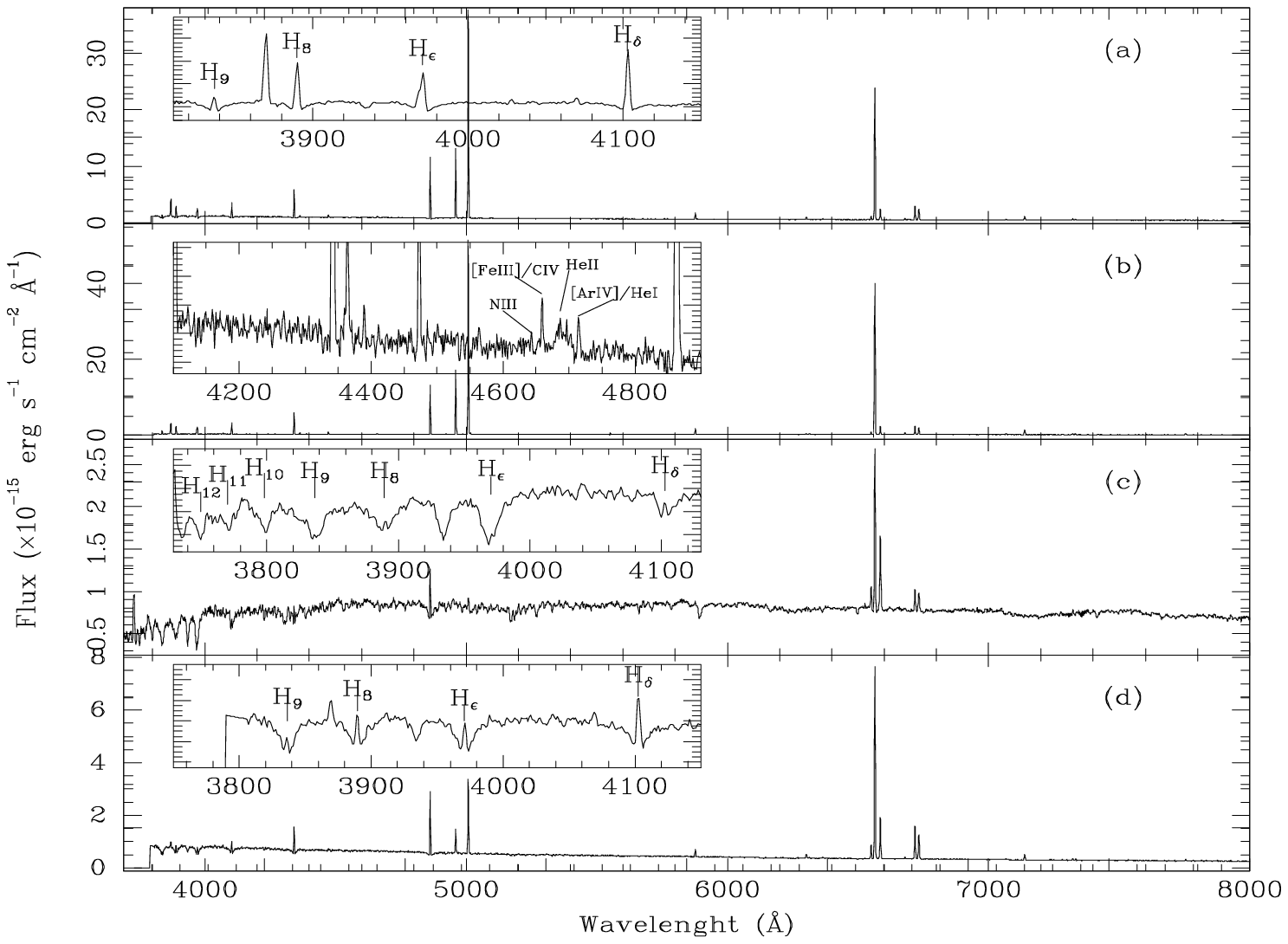} \caption{The optical
spectra of the starburst galaxies in our sample, taken from the
Sloan Digital Sky Survey. The insets of (a), (c) and (d) show
enlargement of the wavelength range of 3750 - 4150 \AA, where the
higher-order Balmer absorption lines are positively detected. The
inset of (b) shows the magnified view of the W-R bumps around 4650
\AA. (a) A1212+06; (b) A1216+04; (c) CGCG 13-83; (d) IC 225.}
\end{center}
\end{figure}

\subsection{Radial Surface Brightness Distribution}

The typical radial surface brightness distribution for ordinary
elliptical galaxies is the de Vaucouleurs $R^{1/4}$ law (de
Vaucouleurs 1953). Since the SDSS photometrical survey at u- and
z-bands are much shallower(Fukugita et al. 1996) and i-band used
a thinned CCD, an instrumental effect
known as the ``red halo" in the PSF wings are reported to
seriously affected the i-band photometric analysis (e.g., Wu et
al. 2005). We will only derive g- and r-band stellar surface brightness
distributions by using the standard task \emph{ellipse} in \emph{IRAF},
for A1212+06 and CGCG 13-83, which are shown in Fig. 4. The SDSS
magnitude system (Fukugita et al. 1996) is quite similar to the AB
system(Oke \& Gunn 1983). The zero point of magnitude for each
frame is obtained directly from the image header. For A1216+04, we
do not apply the photometric measurement since it is a late-type
blue compact dwarf (BCD) galaxy (Gordon \& Gottesman 1981; Hoffman
et al. 1987). The best fits with the de Vaucouleurs $R^{1/4}$ are shown as solid
lines, fitting results are summarized in Table 3. We find that the
fitting is quite well, and the rms is typically less than 0.1.

 In order to derive the g-r color distribution, we first check the
 PSF profiles of g- and r-band images and then smooth the g-band
 image by convolving a Gaussian kernel to match r-band PSF. In
 Figure 5 we show the g-r color distributions for A1212+06(left)
 and CGCG13-83(right). It is very interesting to note that two
 sources show different behaviors. For the case of A1212+06, the
 nucleus is exceptionally blue, and becomes much redder
 continuously outwards. While for CGCG13-83, the color
 distribution is nearly uniform, very similar to a compact
 elliptical galaxy IC 1639 (Wu et al. 2005) except for the inner
 $\sim$5 arcseconds region. Our results indicate their modes
 of star forming activity might by different.

\begin{figure}
\begin{center} \epsscale{1.0}{1.0} \plotone{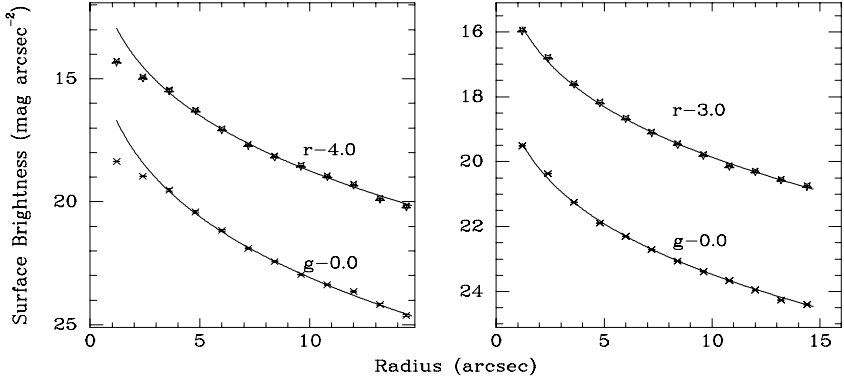} \caption{The stellar
surface brightness of  2-band (gr) SDSS images, the solid lines
are the best fit with de Vaucouleurs R$^{1/4}$ law. The profiles
are shifted for clarity. \emph{Left panel:} A1212+06; \emph{Right
panel:} CGCG 13-83.}
\end{center}
\end{figure}

\begin{figure}
\begin{center}\plotone{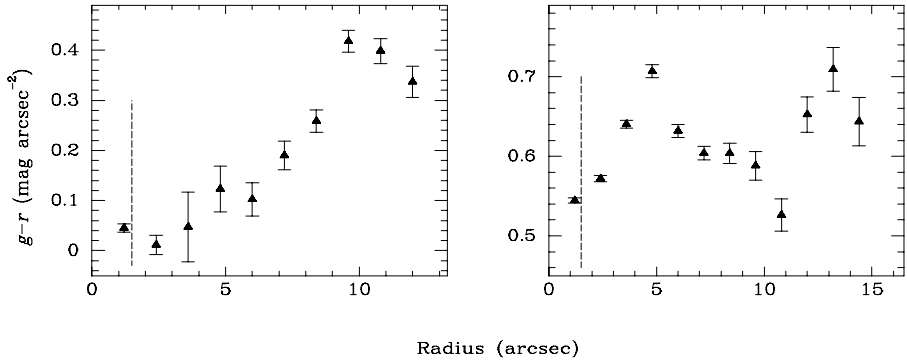} \caption{The
color (g-r) distributions for A1212+06 (left) and CGCG 13-83 (right).
The vertical dashed line indicates the SDSS spectral fibre size.}
\end{center}
\end{figure}

\begin{table}[h]
\caption{The fitting results with de Vaucouleurs R$^{1/4}$ law for
A1212+06 and CGCG 13-83.}
\begin{center}
\begin{tabular}{lcccccc}
\hline\noalign{\smallskip}
&&A1212+06& & &CGCG 13-83&\\
\hline\noalign{\smallskip}
Band & $\mu_0$ &r$_e$ &rms & $\mu_0$ &r$_e$ &rms\\
     & (mag arcsec$^{-2}$)   & (arcsec)&  & (mag arcsec$^{-2}$)   & (arcsec)& \\
\hline\noalign{\smallskip}
g& 7.6$\pm$0.3 & 0.83$\pm$0.06  & 0.07 &13.6$\pm$0.1 &5.2$\pm$0.3&0.06 \\
r& 8.8$\pm$0.4 & 1.2$\pm$0.1  & 0.09 &13.1$\pm$0.1 &5.3$\pm$0.3&0.07 \\
\hline\noalign{\smallskip}
\end{tabular}
\end{center}
\end{table}

\subsection{Star Forming Activity}

\subsubsection{A1212+06}
 A1212+06 is one member of the Virgo cluster and was identified as an
 HII galaxy by Maza et al. (1991) using the emission line ratios
 such as [O II]$\lambda 3727$/[O III]$\lambda 5007$, [N II]$\lambda
 6584$/H$\alpha$. Our stellar population synthesis fitting indicates
 that the monochromatic contribution at 4800\AA \ from the young
 ($<$ 10$^8$ yr), intermediate-aged (10$^8$ $<$ age $<$10$^9$ yr)
 and  old ($>$ 10$^9$ yr) stellar populations are 50\%, 38\% and
 12\%, respectively.

 We use the standard method to compute the nebular extinction,
 based on the Balmer decrement and assuming Case B recombination
 and a standard reddening law (Cardelli, Clayton \& Mathis 1989).
 For A1212+06, the observed $F_{H\alpha}/F_{H\beta}$ is 2.94, the
 nebular extinction, $A_{V}$, is estimated to be 0.076 mag. Thus
 the extinction-corrected H$\alpha$ luminosity,
 $L_{H\alpha}^{corr}$, is 9.3 $\times 10^{39} \ erg \ s^{-1}$.
 Using the empirical calibration given by Kennicutt (1998),
 SFR$_{H\alpha}$ is equal to 0.073 $M_\odot \ yr^{-1}$, note that
 the SFR$_{H\alpha}$ only accounts for the central 3 arcseconds
 region. At the same time, we could also estimate the SFR from the
 infrared (8-1000 $\mu$m) luminosity (Kennicutt 1998). The IR
 luminosity, L$_{8-1000\mu m}$, which is calculated by using the
 fluxes taken from the IRAS Faint Source Catalog (Moshir et al.
 1989), is equal to $9.11 \times 10^{42} \ erg \ s^{-1}$ and  the
 corresponding SFR$_{IR}$ is 0.41 $M_\odot \ yr^{-1}$, much larger
 than SFR$_{H\alpha}$, which suggests that star-forming activity
 arises from larger area than the SDSS spectral fibre region,
 and also confirmed by the g-r color distribution.

\subsubsection{A1216+04}
 Like A1212+06, A1216+04 is an HII galaxy in the Virgo cluster,
 too (Terlevich et al. 1991). Stellar population synthesis fitting
 indicates that a significant contribution from the young stellar
 components which contributes 86\% of the total monochromatic flux
 at 4800\AA. This fraction is much larger than the one (50\%) in
 A1212+06. The observed $F_{H\alpha}/F_{H\beta}$ is equal to 4.38,
 the corresponding nebulae extinction is 1.17 mag, and SFR for the
 central 3 arcseconds region is 0.21 $M_\odot \ yr^{-1}$, it is
 very interesting to note that SFR deduced from IR emission is
 exactly the same as SFR$_{H\alpha}$, which indicates that the
 star-forming activity is very concentrated in the SDSS fibre
 covering region. In fact, this object is wrongly classified as an
 elliptical galaxy (Gordon \& Gottesman 1981; Hoffman et al.
 1987), and will omit from the further analysis.

 The spectrum of A1216+04 shows the broad bumps at 4650 \AA, as
 shown in the inset in Fig. 2b, which are characteristic of W-R
 stars (Conti 1991). The bump at 4650 \AA\ contains stellar and
 nebular emission lines including He II $\lambda$4686, [Fe III]
 $\lambda$4658, N III $\lambda$$\lambda$4634-4641, [Ar IV]
 $\lambda$4711 and C IV $\lambda$4658. Thus we identify A1216+04
 as a hitherto unknown Wolf-Rayet galaxy, however, the low S/N for
 the Wolf-Rayet features hampers us for further analysis.

\subsubsection{CGCG 13-83}
 CGCG 13-83 has been regarded as an actively star-forming elliptical
 galaxy by Fukugita et al. (2004), who firstly found its strong
 H$\alpha$ emission line. But the authors did not do any further
 analysis such as the stellar population synthesis and the radial
 profiles, which will be presented here. As shown in Fig. 3c, the
 spectrum for CGCG 13-83 clearly shows higher-order Balmer absorption lines
 in the wavelength range of 3750 $-$ 4150 \AA, which have
 been taken as the unambiguous evidence of intermediate-aged ($\sim
 10^8$ yr) stellar populations (Gonzalez Delgado, Leitherer \&
 Heckman 1999; Gonzalez Delgado, Heckman \& Leitherer 2001).
 Our stellar population synthesis modeling confirms that
 46\% of the monochromatic flux at 4800\AA \ is due to the intermediate-aged
 (10$^8$ $<$ age $<$ 10$^9$ yr) stellar components.

 The observed $F_{H\alpha}/F_{H\beta}$ is equal to 4.25, the
 corresponding nebulae extinction is 1.09 mag, and SFR for the
 central 3 arcseconds region is 0.07 $M_\odot \ yr^{-1}$.
 All results are summarized as in Table 4.

 Unlike A1212+06 and A1216+04, the [O III]$\lambda\lambda$4959, 5007
 in CGCG 13-83 is rather weak and it can only be detectable after subtracting
 the best matched model from the observed spectrum.
 However, the radial profiles can be fitted with the R$^{1/4}$ law very well for g-
 and r-band images, as shown in the right panel in Fig. 4.

\begin{table}[h]
\caption{The extinction for the nebular, the extinction-corrected
H$_\alpha$ luminosity and the star formation rate of the three
star-forming galaxies}
\begin{center}
\begin{tabular}{lccccc}
\hline\noalign{\smallskip}
Name & $A_{V}^{nebular}$ & $L_{\rm H\alpha}^{\rm corr}$ &SFR ($H_{\alpha}$) & $L_{\rm FIR}$ &SFR (FIR)\\
     & (mag)             & (erg s$^{-1}$)          &($M_\odot$ yr$^{-1}$) & (erg s$^{-1}$) & ($M_\odot$ yr$^{-1}$)\\
\hline\noalign{\smallskip}
A1212+06  & 0.076 & $9.30 \times 10^{39}$  & 0.073  &$9.11 \times 10^{42}$& 0.41\\
A1216+04  & 1.17  & $2.71 \times 10^{40}$  & 0.21  &$4.60 \times 10^{42}$&0.21\\
CGCG 13-83 & 1.09  & $8.83 \times 10^{39}$  & 0.070&...&...\\
\hline\noalign{\smallskip}
\end{tabular}
\end{center}
\end{table}

\section{Discussion}

 By using images and spectra from the SDSS, we have investigated
 the emission line properties for a sample of 11 elliptical
 galaxies with emission lines picked out from 48 ellipticals by
 cross-correlating the SDSS DR2 with RC3 catalog. We classify four
 HII galaxies, one of which (A1216+04) is in fact a late-type BCD
 and a new Wolf-Rayet galaxy, two (A1212+06 and IC 225) are dwarf
 ellipticals, and one (CGCG13-83) is an ordinary elliptical.
 Recently Guzman et al. (2003) and Graham (2005) find that dwarf
 ellipticals form a continuous extension, both chemically and
 dynamically, with the more luminous (ordinary) ellipticals. If we
 simply regard A1212+06 and IC 225 as normal ellipticals, for
 our sample, the frequency of star-forming ellipticals is $3/47 =
 6.38\%$, significantly higher than the value $2/210\approx1\%$ by
 Fukugita et al. (2004), which is probably caused by our selection
 criteria, small sample size and problems of morphological
 classification in RC3 (de Souza, Gadotti \& dos Anjos 2004).

 It is well known that bar, galaxy-galaxy interaction and merger
 are efficient mechanisms to trigger starburst activities for
 spiral galaxies (Huang et al. 1996; Zou et al. 1995).   Since the
 star-forming elliptical galaxies are rather rare, the most
 possible mechanism for triggering starburst in the elliptical
 galaxies is still an open question. The most promising mechanisms
 to trigger the nuclear starbursts in the elliptical galaxies
 include interaction, major mergers and minor mergers (accretion
 events with the satellite less than 10\% of the galaxy's mass) as
 suggested by Worthey (1997). Major mergers are the mergers of two
 disc galaxies of comparable mass, using the numerical
 simulations, Mihos \& Hernquist (1996) have shown that starburst
 activities during the merger process could be two orders of
 magnitude higher than that in isolated galaxies and can be
 sustained from several 10$^7$ yr to $\sim$ 2 $\times$ 10$^8$ yr
 after the collision. It could be the case of many infrared
 luminous galaxies, which are found to show morphological
 peculiarities indicative of encounters, such as multiple nuclei,
 tidal tails, loops, and shells (Sanders et al. 1988; Sanders
 1992). Though major mergers trigger the most powerful starburst,
 they are less common than minor mergers with the satellite less
 than 10\% of the galaxy's mass. Ostriker \& Tremaine (1975) and
 Tremaine (1981) have shown that for the typical galaxy, no matter
 of morphological type, several tens of percents of its mass has
 probably been accreted in the form of discrete subunits. And the
 numerical simulation of minor mergers between gas-rich disk
 galaxies and less massive dwarf galaxies are shown by Hernquist
 \& Mihos (1995).

 For CGCG 13-83, the light distribution is very smooth, there is
 no any sign of interacting remnant, such as tidal tails. However,
 when we search in the view field of 30 arcminutes by using the
 NASA/IPAC Extragalactic Database (NED), we detect two galaxies
 that possibly interact with CGCG 13-83. One is CGCG 13-84 (RA:
 12h08m31.3s, Dec: +00d08m11s; z: 0.034942), which is about 2.5
 arcminutes northeast to CGCG 13-83 and 15.2 mag (g-band). The
 other is SDSS J120828.48+000948.7 (RA: 12h08m28.490s, Dec:
 +00d09m48.90s; z: 0.03539), which is about 3.4 arcminutes
 northwest to CGCG 13-83 and 17.3 mag (g-band). Therefore, CGCG
 13-84 is a most probably physical companion to CGCG13-83
 according to the criteria proposed by Schmitt (2001). Meanwhile,
 previous numerical simulations indicate that gas inflows occur
 within a dynamical timescale (a few $\times$ 10$^8$ yr) of the
 initial collision (Barnes \& Hernquist 1991; Mihos et al. 1992,
 1993), both suggest that the triggering mechanism for star
 forming activity in CGCG 13-83 is due to the interaction with
 CGCG 13-84.

 For the case of A1212+06, the morphology is also very smooth.
 However, when we inspect the central region carefully,  we find
 two distinct cores, which are clearly shown in its RGB image. Due
 to the no-spatial information of SDSS spectrum, now we can't
 study the nature for both cores. For this object, long-slit
 optical spectroscopy is under consideration.

\section{Conclusions}
 In this paper, we study a sample of 11 elliptical galaxies with
 emission lines. After removing old stellar contribution and using
 the standard classification criteria, we identify them into one
 Seyfert 2, six LINERs and four HII galaxies. For four HII
 galaxies, we find that one (A1216+04) is a new Wolf-Rayet galaxy,
 two ellipticals (A1212+06, IC 225) have two distinct nucleus, and
 CGCG 13-83 is possibly interacting with CGCG 13-84. We propose
 that the star-forming activities in elliptical galaxies could be
 triggered by either galaxy-galaxy interaction or the merging of a
 small satellite/a massive star cluster.

\begin{acknowledgements}
The authors are very grateful to the anonymous referee for very
carefully reading and constructive comments which significantly
improved the content of the paper. This work is supported by the
National Natural Science Foundation
of China under grants 10103001 and 10221001 and the National Key
Basic Research Science Foundation (NKBRSG19990754). Funding for
the creation and distribution of the SDSS Archive has been
provided by the Alfred P. Sloan Foundation, the Participating
Institutions, the National Aeronautics and Space Administration,
the National Science Foundation, the U.S. Department of Energy,
the Japanese Monbukagakusho, and the Max Planck Society. The SDSS
web site is http://www.sdss.org/.  The SDSS is managed by the
Astrophysical Research Consortium (ARC) for the Participating
Institutions. The Participating Institutions are The University of
Chicago, Fermilab, the Institute for Advanced Study, the Japan
Participation Group, The Johns Hopkins University, Los Alamos
National Laboratory, the Max-Planck-Institute for Astronomy
(MPIA), the Max-Planck-Institute for Astrophysics (MPA), New
Mexico State University, University of Pittsburgh, Princeton
University, the United States Naval Observatory, and the
University of Washington. This research has made use of the
 NASA/IPAC Extragalactic Database (NED) which is operated by the
 Jet Propulsion Laboratory, California Institute of Technology,
 under contract with the National Aeronautics and Space
 Administration.
\end{acknowledgements}

\end{document}